\documentstyle[preprint,pra,aps]{revtex}

\begin{document}

\draft

\tightenlines

\title{Theory of dressed states in quantum optics}

\author{Marco Frasca}
\address{Via Erasmo Gattamelata, 3,
         00176 Roma (Italy)}

\date{\today}

\maketitle

\abstract{
The dual Dyson series [M.Frasca, Phys. Rev. A
{\bf 58}, 3439 (1998)], is used to develop a general perturbative method
for the study of atom-field interaction in quantum optics. In fact,
both Dyson series and its dual, through renormalization group
methods to remove secular terms from the perturbation series, give
the opportunity of a full study of the solution of the Schr\"{o}dinger
equation in different ranges of the parameters of the given hamiltonian.
In view of recent experiments with strong laser fields, this approach
seems well-suited to give a clarification and an improvement
of the applications of the dressed states as currently
done through the eigenstates of the atom-field interaction,
showing that these are just the leading order of the
dual Dyson series when the Hamiltonian is expressed in the
interaction picture. In order to exploit the method at the best,
a study is accomplished
of the well-known Jaynes-Cummings model in the rotating
wave approximation, whose exact solution is known,
comparing the perturbative solutions
obtained by the Dyson series and its dual
with the same approximations obtained by Taylor
expanding the exact solution.
Finally, a full perturbative study of high-order harmonic
generation is given obtaining, through analytical expressions,
a clear account of the power spectrum
using a two-level model, even if the method can be successfully
applied to a more general model that can account for ionization
too. The analysis shows that to account for the power spectrum
it is needed to go to first order in the perturbative analysis.
The spectrum obtained gives a way to measure experimentally the
shift of the energy levels of the atom interacting with the
laser field by looking at the shifting of hyper-Raman lines.
}


\narrowtext

\section{Introduction}

Recent experiments on atoms using strong laser
fields \cite{exp} have shown the appearance of
a wealth of new effects, e.g. high-order harmonics generation,
in the interaction between light and atoms.
This situation forced researchers to find different approaches
to describe the outcomes of those experiments.
Numerical studies of the time-dependent
Schr\"{o}dinger equation \cite{ebe} have shown that
the two-level model still proves to be very
useful to describe all the features of
harmonics generation\cite{mil},
even if the rotating wave approximation
must be abandoned. Indeed, recent work \cite{kei,maq}
indicates, by comparing results from a two-level
model using Floquet states and numerical work
on the Schr\"{o}dinger equation, that the simple
two-level model is fairly effective in describing
the physical situation at hand. So far, no
perturbative solution seems to be known of
this two-level model beyond Floquet states
for the case of a strong laser field.
But, a study by Meystre of an atom in
a Fabry-Perot cavity \cite{mey} used the
same model of Ref.\cite{maq} and gave
a first perturbative analytical solution
to such a model in a strong coupling regime.
In fact, the analytical solution given by Meystre
and its higher order corrections has
been successfully obtained in Ref.\cite{rg},
showing that the levels of the atom undergoes
a shift. Being the same model, now
we have at hand a way to observe experimentally
such a shift through hyper-Raman lines in
harmonic generation, if one is able to properly
account for the spectrum.

An understanding of interaction between an atom and
a strong electromagnetic field has been possible in recent years
through the introduction of the dressed-atom picture\cite{scul}.
This approach assumes that the field couples the levels of the
atom in such a way that the interaction is between this ``dressed''
atom and the field itself.
The computation of the corresponding dressed states,
as currently found in literature,
involves or the computation of the eigenstates and the eigenvalues
of the term of interaction
between the atom and the field in the Hamiltonian either the computation
of the eigenstates of the full Hamiltonian, taking in this way
into account the field too. From a physical standpoint the dressed-atom
picture is quite general as it assumes that the photons
of the field surround the atom as to modify the way the atom itself
responds to the field, then it should concern a fully
second quantized theory. But, the computation of the
eigenstates of the full Hamiltonian or just the atom-field interaction term,
that we take to be the dressed states, often reveals itself as
an approximation scheme whose understanding is the main aim of this paper.
So far, no reason has been known for the nice working of such dressed
states in applied mathematics. A recently
devised approach \cite{fra1}, the dual Dyson perturbation
series, turns out to be both an explanation and an improvement
of the computation of dressed states permitting the
computation of higher order corrections to a leading
order solution obtained through such dressed states. As a
by-product one has a clear physical understanding of
what are the parameters involved in such approximate dressed states
and what is going to neglect. So, by this improvement
of the computation of dressed states, we are able to find an analytical
perturbative solution to the two-level model to
analyse high-order harmonic generation showing that
this is a first order effect, that is, the leading order
solution found by Meystre
is not enough to get the right spectrum. Then, the
result properly accounts for the relevance of population
distribution as discussed in Ref.\cite{maq} and an
analytical closed expression is given.

The dual Dyson series that accounts for the dressed states
as defined above
can be derived from the time-dependent Schr\"{o}dinger equation by
using the duality principle in perturbation
theory and the quantum
adiabatic approximation \cite{fra1}.
In this way one realizes that the dual Dyson series
is the same one of Ref.\cite{most}. The results
one gets from what should work just for quantum adiabatic
processes can appear somewhat unexpected, as it will be
shown for the Jaynes-Cummings model in the
rotating wave approximation (RWA) for whom
an exact solution is known. But, this
just agrees with the results of Ref.\cite{fra1}.

So, the existence of a dual Dyson series can improve
the study of atom-field interaction. In fact, one
can accomplish a perturbative analysis of models in
quantum optics in different regions of the parameter
space that for a Jaynes-Cummings model can be easily
identified, when spontaneous emission is neglected,
with the ratio between the detuning and the Rabi
frequency. Then, by generalizing the computation of
dressed states through the dual Dyson series on one side
and by the standard Dyson series on the other, we can reach the
main aim of this paper:
A general perturbative method to study atom-field
interaction in quantum optics at different values
of the parameters of the Hamiltonian.

The completeness of our approach is strongly tied
with the recent results obtained in quantum optics
through the renormalization group methods for
perturbation theory \cite{rg}. These methods
permit the resummation of the so called secularities
that appear in perturbation theory.
Indeed, we are able to derive a energy level shift of
the atom in high-order harmonic generation that
has effect on hyper-Raman lines. As shown in Ref.\cite{maq},
when the two levels of the atom are equally populated,
only hyper-Raman lines should be observed. Then, in
view of this situation, such an energy level shift
turns out to be significant.

It should be pointed out that, although
the extension of this approach
to the method of the master equation \cite{scul} should
be straigthforward, it is not considered in this paper.
So, e.g. the effect of vacuum fluctuations of the field
modes is neglected.

The paper is so structured. In sec.II we give a general
description of the methods and show why the
eigenstates of the perturbation
are important for strong fields. In sec.III
a study of the Jaynes-Cummings model in RWA
is accomplished in order to have
a pedagogical description of the methods and a
comparation with an exact solution. In sec.IV
the question of high-order harmonic generation
is discussed through the methods so far introduced.

\section{A general method for perturbative analysis}

\subsection{General theory}

In Ref.\cite{fra1} we have introduced the duality principle
in perturbation theory. By duality we mean that, for a given
differential equation, it is possible to compute both a
perturbation series in $\lambda$ and $\frac{1}{\lambda}$,
being $\lambda$ the characteristic parameter of the equation.
This is accomplished by a proper choice of the leading
order equation. So, e.g. for the Duffing equation
\begin{equation}
    \ddot{x}+x+\lambda x^3=0
\end{equation}
one can compute a series in $\lambda$ and $\frac{1}{\lambda}$
by taking, at leading order, in the former case
$\ddot{x}+x=0$ and in the latter case $\ddot{x}+\lambda x^3=0$.
It easy to see that the duality principle
is true indipendently by our ability to do the computations
of the equations one gets from the perturbation series.

In turn, the existence of a duality principle in perturbation
theory means that is possible a perturbative analysis in
different regions of the parameter space of the given
equation. This situation could turn out to be very useful in quantum
mechanics if one is able to obtain a dual Dyson series. This
is indeed the case.

So, let us consider the time-dependent Schr\"{o}dinger equation
\begin{equation}
    H(t)|\psi\rangle=i\frac{\partial |\psi\rangle}{\partial t}
\end{equation}
being $H(t)$ the Hamiltonian and $\hbar=1$ here and in the following.
The Dyson series is a perturbative solution of this equation given by
\begin{equation}
    |\psi(t)\rangle = \left(I-i\int_{t_0}^t dt_1 H(t_1)
    -\int_{t_0}^t dt_1\int_{t_0}^{t_1} dt_2 H(t_1) H(t_2)+\cdots\right)
    |\psi(t_0)\rangle \label{eq:dyson}
\end{equation}
or, by introducing the time ordering operator ${\cal T}$,
\begin{equation}
    |\psi(t)\rangle =
    {\cal T}\exp\left(-i\int_{t_0}^t dt'H(t')\right)|\psi(t_0)\rangle.
\end{equation}
The dual series can be obtained, through the duality principle, by
assuming that the Hamiltonian $H(t)$ has a discrete spectrum, that is
$H(t)|n,t\rangle=E_n(t)|n,t\rangle$ with $|n,t\rangle$ the
eigenstate corresponding to the eigenvalue $E_n(t)$. Then, the
dual Dyson series is the one given in Ref.\cite{most}, that is
\begin{equation}
    |\psi(t)\rangle = U_A(t)
    {\cal T}\exp\left(-i\int_{t_0}^t d\hat{t}
    H'(\hat{t})\right)|\psi(t_0)\rangle   \label{eq:ddyson}
\end{equation}
being
\begin{equation}
    U_A(t)=\sum_n e^{i\gamma_n(t)-i\int_{t_0}^t dt'E_n(t')}
    |n,t\rangle\langle n,t_0|  \label{eq:ua}
\end{equation}
the adiabatic unitary evolution operator, for the Berry phase
$\dot{\gamma}_n(t)=\langle n,t|
i\frac{\partial}{\partial t}|n,t\rangle$ and
\begin{equation}
    H'(t)=
    -\sum_{n,m,n\neq m}
    e^{-i(\gamma_m(t)-\gamma_n(t))}
    e^{i\int_{t_0}^tdt'(E_m(t')-E_n(t'))}
    \langle m,t|i\hbar\frac{\partial}{\partial t}|n,t\rangle
    |m,t_0\rangle\langle n,t_0|. \label{eq:h1}
\end{equation}
This result proves that the well-known adiabatic approximation
and its higher order corrections
can be very effective in building asymptotic approximations to
the solution of the Schr\"{o}dinger equation, as is, on the
other side, the Dyson series.

Let us now consider a perturbed quantum system with Hamiltonian
\begin{equation}
    H=H_0+V(t)
\end{equation}
being $H_0$ the Hamiltonian of the unperturbed system and
$V(t)$ the perturbation. In the interaction picture one has
\begin{equation}
    H_I(t)=e^{iH_0t}V(t)e^{-iH_0t}.
\end{equation}
It is now possible to study the given system
in different regions of the parameter
space  through the Dyson series and its dual.
In the former case we have standard textbook
time-dependent perturbation theory.
In the latter case we have to compute
\begin{equation}
    H_I(t)|n,t\rangle_I=E^{(I)}_n(t)|n,t\rangle_I.
\end{equation}
But $H_I(t)$ is just the interaction $V(t)$ transformed
by an unitary transformation.Then, the eigenvalues $E^{(I)}_n(t)$
are those of the perturbation $V(t)$ and the eigenstates $|n,t\rangle_I$
are just an unitary transformation away from the corresponding eigenstates.
These are the dressed states as generally computed in the current
literature: It is just the leading order approximation of a
dual Dyson series. But now we have a more
general theory and higher order corrections can be computed.
Beside, we realize why the dressed states are so effective
in a strong field regime being obtained from the dual
Dyson series that has a development parameter
exactly inverse of the one of the Dyson series.

It should be pointed out that both Dyson series and
its dual can have the same kind of problems. One
of the most important is surely the question of
secularities: In any case,
resummation of secular terms can be achieved
through the renormalization group methods as
pointed out, for quantum optics, in Ref.\cite{rg}.

\subsection{An example}

To give a clear insight of the working of the above analysis
for a differential equation, let us consider the standard
textbook example
\begin{equation}
    \psi''(x)+\alpha^2(x)\psi(x)=0.
\end{equation}
that can be written in the form (the $i$ factor
is introduced just for convenience)
\begin{equation}
    i\frac{d}{dx}
    \left(\begin{array}{c} \psi(x) \\ \phi(x) \end{array}\right)
    =
    \left(\begin{array}{clcr}
          0 & i \\
          -i\alpha^2(x) & 0
          \end{array}\right)
    \left(\begin{array}{c} \psi(x) \\ \phi(x) \end{array}\right)
    =L(x)\left(\begin{array}{c} \psi(x) \\ \phi(x) \end{array}\right).
\end{equation}

We can apply Dyson series and its dual. Dyson series is not
normally applied to the above equation. Indeed, it gives the
expansion
\begin{equation}
    \left(\begin{array}{c} \psi(x) \\ \phi(x) \end{array}\right)
    = \left[I-i\int_{x_0}^x dx'
    \left(\begin{array}{clcr}
          0 & i \\
          -i\alpha^2(x') & 0
          \end{array}\right)
          -\int_{x_0}^{x}dx'\int_{x_0}^{x'}dx''
    \left(\begin{array}{clcr}
          \alpha^2(x'') & 0 \\
           0 & \alpha^2(x')
          \end{array}\right)+\cdots\right]
    \left(\begin{array}{c} \psi(x_0) \\ \phi(x_0) \end{array}\right).
\end{equation}

In order to compute the dual Dyson series, we need to compute
the eigenvectors and eigenvalues of the matrix $L(x)$. So, for
the eigenvalue $\alpha(x)$ one has the eigenvector
\begin{equation}
    |1,x\rangle=\frac{1}{\sqrt{-2i\alpha(x)}}
    \left(\begin{array}{c} 1 \\ -i\alpha(x) \end{array}\right)
\end{equation}
and for the eigenvalue $-\alpha(x)$
\begin{equation}
    |2,x\rangle=\frac{1}{\sqrt{2i\alpha(x)}}
    \left(\begin{array}{c} 1 \\ i\alpha(x) \end{array}\right).
\end{equation}
Then, one has for the Berry phases $\langle 2,x|i\frac{d}{dx}|2,x\rangle
=\langle 1,x|i\frac{d}{dx}|1,x\rangle=0$ and the
unitary evolution operator (\ref{eq:ua})
\begin{equation}
    U_A(x,x_0)=\frac{1}{\sqrt{\alpha(x)\alpha(x_0)}}
    \left(\begin{array}{clcr}
          \alpha(x_0)\cos\left(\int_{x_0}^xdx'\alpha(x')\right)
           &  \sin\left(\int_{x_0}^xdx'\alpha(x')\right) \\
           -\alpha(x)\alpha(x_0)\sin\left(\int_{x_0}^xdx'\alpha(x')\right)
           & \alpha(x)\cos\left(\int_{x_0}^xdx'\alpha(x')\right)
          \end{array}\right)
\end{equation}
It is straightforward to see that
\begin{equation}
\left(\begin{array}{c} \psi(x) \\ \phi(x) \end{array}\right)\approx
    U_A(x,x_0)
\left(\begin{array}{c} \psi(x_0) \\ \phi(x_0) \end{array}\right)
\end{equation}
gives the well-known Wentzel-Kramers-Brillouin-Jeffreys (WKBJ) result
\begin{equation}
   \psi(x) \approx \frac{C_1}{\sqrt{\alpha(x)}}
   \cos\left(\int_{x_0}^xdx'\alpha(x')\right)+
   \frac{C_2}{\sqrt{\alpha(x)}}
   \sin\left(\int_{x_0}^xdx'\alpha(x')\right).
\end{equation}
In this derivation we have omitted the problem connected to
turning points. We just note that, if there are points where
$\alpha(x)=0$, Berry phases are no more zero as these
are degeneracy points.

This example shows the full power of the adiabatic approximation in
finding asymptotic approximations to a given differential equation,
without any requirement of slowly variation of the parameters of the equation.
In the following we will show how to find higher order corrections too.

\subsection{Duality and Berry's asymptotics}

Duality principle has been introduced in Ref.\cite{fra1} to resolve
problems both with infinitely small and large perturbations. As
such, there is a region of the parameter space that is not possible
to analyse by perturbation methods.
But, it is not difficult to realize that,
as a by-product, an alternative solution to the Schr\"{o}dinger
equation for its unitary evolution through eq.(\ref{eq:ddyson})
is obtained. This has no trivial consequences as, differently
from the Dyson series, a superadiabatic scheme could be applied
instead, as devised by Berry \cite{Berry} that could give
non-perturbative informations on the dual series.

A superadiabatic scheme proves to be very useful when the full
Hamiltonian is considered with no a priori large or small parts,
as shown in Ref.\cite{Holth} to describe stimulated Raman adiabatic
passage by a three-level model. Indeed, the idea is
to iterate the scheme to compute the adiabatic series giving
$U_A(t)$ and $H'(t)$, by computing $U'_A(t)$ for $H'(t)$, and the
new Hamiltonian $H''(t)$ through the eigenstates of $H'(t)$.
In principle, the procedure can be repeated to the step one
wants, giving the unitary evolution
$U(t)\sim U_A(t)U'_A(t)U''_A(t)\cdots U^{(n)}(t)$ and it is
tempting to stop to a given step to obtain an approximation
to the unitary evolution but, actually, the procedure
is shown to diverge. Anyhow, an optimal step $n_c$ exists for
which an eigenstate basis set can be build by the approximated $U(t)$
to approximate the solution of the Schr\"{o}dinger equation.
Divergence is due to the fact that off-diagonal terms
computed by the new Hamiltonians are systematically neglected.

Indeed, to address the question of dressed states
we consider a Hamiltonian like
\begin{equation}
    H=\frac{\omega_0}{2}\sigma_3 + V(t)\sigma_1
\end{equation}
being $V(t)$ a generic perturbation, $\sigma_1$ and $\sigma_3$ Pauli
matrices and $\omega_0$ the level separation of the model. The regimes
of interest are fully perturbative as $V(t)$ is assumed to be very large. So,
the initial Hamiltonian to apply the superadiabatic scheme is given, in
interaction picture, by
\begin{equation}
    H_I=e^{i\omega_0 t\sigma_3}V(t)\sigma_1.
\end{equation}
In this case, the superadiabatic scheme just stop to the second step
as, at first step one has $U_A(t)=e^{i\frac{\omega_0}{2}\sigma_3 t}
e^{-i\sigma_1\int_0^tdt'V(t')}$ and, at the second step,
$U'_A(t)=U^\dagger_A(t)$, so the product of unitary evolution
operators is stopped and nothing new is obtained. Anyhow, the Berry's
scheme can prove to be very useful in a non-perturbative regime, that
is, when $V(t)$ and $\omega_0$ are of the same order of magnitude and
exponentially small factors can be retained. Then,
we can conclude that a superadiabatic scheme turns out to be useful in an
intermediate regime, being in this way a bridge between the small and large
perturbation theory linked in turn by the duality principle. This matter
deserves further investigation.

\section{Perturbative analysis of the Jaynes-Cummings model}

The Jaynes-Cummings model is widely used in quantum optics. Its
Hamiltonian, in the RWA, is given by \cite{scul}
\begin{equation}
    H_{JC}=\omega a^\dagger a
    +\frac{\omega_0}{2}(|2\rangle\langle 2|-|1\rangle\langle 1|)
    +g(|2\rangle\langle 1|a^\dagger +|1\rangle\langle 2|a)
\end{equation}
representing a two-level atom coupled with a single mode radiation of
frequency $\omega$ through the constant $g$. The reason to
consider it here is that the exact solution is known and can be
compared with the results of our perturbative analysis.

In the interaction picture one has the Hamiltonian
\begin{equation}
    H_{JC}^{(I)}=g(e^{i\Delta t}|2\rangle\langle 1|a^\dagger +
    e^{-i\Delta t}|1\rangle\langle 2|a)
\end{equation}
being $\Delta=\omega_0-\omega$ the detuning that here we
assume different from $0$ for the sake of generality. As
it can be seen from the form of $H_{JC}^{(I)}$, the
critical parameter in the model is the ratio
$\frac{g}{\Delta}$. This means that an eventual
perturbation series and its dual
will have this parameter and its inverse
as a development parameter. Now, we proceed to compute
those series from the exact solution.

The exact solution of the Schr\"{o}dinger equation in
interaction picture
\begin{equation}
    H_{JC}^{(I)}|\psi\rangle_I=i\frac{\partial|\psi\rangle_I}{\partial t}
\end{equation}
can be found by looking for a solution in the form
\begin{equation}
    |\psi\rangle_I=\sum_n c_{1,n+1}(t)|1,n+1\rangle+c_{2,n}(t)|2,n\rangle
\end{equation}
being $n$ the photon number.
So, the probability amplitudes are given by \cite{scul}
\begin{eqnarray}
    c_{1,n+1}(t)&=&\left\{c_{1,n+1}(0)\left[
    \cos\left(\frac{\Omega_nt}{2}\right)+\frac{i\Delta}{\Omega_n}
    \sin\left(\frac{\Omega_nt}{2}\right)\right]-\frac{2ig\sqrt{n+1}}
    {\Omega_n}c_{2,n}(0)\sin\left(\frac{\Omega_nt}{2}\right)
    \right\}e^{-i\Delta t/2} \nonumber \\
    & & \label{eq:sjc} \\
    c_{2,n}(t)&=&\left\{c_{2,n}(0)\left[
    \cos\left(\frac{\Omega_nt}{2}\right)-\frac{i\Delta}{\Omega_n}
    \sin\left(\frac{\Omega_nt}{2}\right)\right]-\frac{2ig\sqrt{n+1}}
    {\Omega_n}c_{1,n+1}(0)\sin\left(\frac{\Omega_nt}{2}\right)
    \right\}e^{i\Delta t/2} \nonumber
\end{eqnarray}
being $\Omega_n=\sqrt{\Delta^2+{\cal R}^2_n}$ and
${\cal R}_n=2g\sqrt{n+1}$ the Rabi frequency. As expected, being
$\Delta$ and $g$ the only parameters, their ratio enters
the only meaningful development parameter. The Dyson series
is obtained by expanding the above solution in Taylor series of
$\lambda=\frac{{\cal R}_n}{\Delta}$ giving till second order
\begin{eqnarray}
    c_{1,n+1}(t)&=&\left\{c_{1,n+1}(0)\left[
    1+
    i\frac{\lambda^2}{4}
    \left(\Delta t +
    i(1-e^{-i\Delta t})\right)
    \right]
    -\frac{\lambda}
    {2}c_{2,n}(0)(1-e^{-i\Delta t})+O(\lambda^3)
    \right\} \nonumber \\
    & & \label{eq:sjca} \\
    c_{2,n}(t)&=&\left\{c_{2,n}(0)\left[
    1-i\frac{\lambda^2}{4}
    \left(\Delta t +i
    (e^{i\Delta t}-1)\right)
    \right]
    -\frac{\lambda}
    {2}c_{1,n+1}(0)(e^{i\Delta t}-1)+O(\lambda^3)
    \right\}. \nonumber
\end{eqnarray}
It is easy to see that at second order in the development parameter a
secularity appears, that is a term that grows without bound in the
limit $t\rightarrow\infty$. In perturbation theory,
unless we are not able to get
rid of the secularity the series is not very useful.
This can be accomplished through the renormalization group
methods described in Ref.\cite{rg}. But here, the problem can be
easily traced back to the Taylor expansion of the functions
$\sin(\sqrt{1+\epsilon^2}t)$
in $\epsilon$, having $\sqrt{1+\epsilon^2}=1+\frac{\epsilon^2}{2}+
O(\epsilon^4)$. So, we can eliminate it
by simply substituting $\Delta$ with $\Delta + \frac{{\cal R}_n^2}{2\Delta}$
everywhere in the approximate solution into the exponentials
of eq.(\ref{eq:sjca}).

It is not difficult to get back the result (\ref{eq:sjca}) through
the Dyson series (\ref{eq:dyson}). So, as expected, this series gives
an analysis of the Jaynes-Cummings model when the detuning $\Delta$ is
enough larger than the Rabi frequency ${\cal R}_n$.

Now, let us repeat the above discussion in the opposite limit
with the Rabi frequency larger than the detuning. Again,
by Taylor expanding the exact solution one has
\begin{eqnarray}
    c_{1,n+1}(t)&=&\left\{c_{1,n+1}(0)\left[
    \cos\left(\frac{{\cal R}_n}{2}t\right)
    +\frac{i}{\lambda}
    \sin\left(\frac{{\cal R}_n}{2}t\right)-
    \frac{1}{2\lambda^2}\frac{{\cal R}_n}{2}t
    \sin\left(\frac{{\cal R}_n}{2}t\right)
    \right]\right. \nonumber \\
    &-&
    \left.ic_{2,n}(0)
    \left[\sin\left(\frac{{\cal R}_n}{2}t\right)
    -\frac{1}{2\lambda^2}
    \left(\sin\left(\frac{{\cal R}_n}{2}t\right)-\frac{{\cal R}_n}{2}t
    \cos\left(\frac{{\cal R}_n}{2}t\right)\right)\right] +
    O\left(\frac{1}{\lambda^3}\right)
    \right\}e^{-i\Delta t/2} \nonumber \\
    & & \label{eq:sjcb} \\
    c_{2,n}(t)&=&\left\{c_{2,n}(0)\left[
    \cos\left(\frac{{\cal R}_n}{2}t\right)-\frac{i}{\lambda}
    \sin\left(\frac{{\cal R}_n}{2}t\right)
    -\frac{1}{2\lambda^2}\frac{{\cal R}_n}{2}t
    \sin\left(\frac{{\cal R}_n}{2}t\right)
    \right]\right. \nonumber \\
    &-& \left.ic_{1,n+1}(0)
    \left[\sin\left(\frac{{\cal R}_n}{2}t\right)
    -\frac{1}{2\lambda^2}
    \left(\sin\left(\frac{{\cal R}_n}{2}t\right)-\frac{{\cal R}_n}{2}t
    \cos\left(\frac{{\cal R}_n}{2}t\right)\right)\right]+
    O\left(\frac{1}{\lambda^3}\right)
    \right\}
    e^{i\Delta t/2} \nonumber
\end{eqnarray}
with the same problem of a secularity at second order. Indeed, this
series can be obtained by the dual Dyson series (\ref{eq:ddyson})
showing what could seem an unexpected result from the adiabatic
approximation, but in agreement with the results of Ref.\cite{fra1}.

To compute the dual Dyson series we need the eigenstates and
eigenvalues of $H_{JC}^{(I)}$. It is easily found that for
the eigenvalue $g\sqrt{n+1}$ we have the eigenstate
\begin{equation}
    |a,n,t\rangle=\frac{1}{\sqrt{2}}
    (e^{-i\Delta t}|1,n+1\rangle+|2,n\rangle)
\end{equation}
and for the eigenvalue $-g\sqrt{n+1}$ we have the eigenstate
\begin{equation}
    |b,n,t\rangle=\frac{1}{\sqrt{2}}
    (|1,n+1\rangle-e^{i\Delta t}|2,n\rangle)
\end{equation}
that are easily recognized as the dressed states
of Ref.\cite{scul} for the
Jaynes-Cummings model with a non-zero detuning. Berry phases are
then easily computed to give
\begin{eqnarray}
    \dot{\gamma}_a &=&
    \langle a,n,t|i\frac{\partial}{\partial t}|a,n,t\rangle =
    \frac{\Delta}{2} \nonumber \\
    & & \\
    \dot{\gamma}_b &=&
    \langle b,n,t|i\frac{\partial}{\partial t}|b,n,t\rangle =
    -\frac{\Delta}{2}.    \nonumber \\
\end{eqnarray}
Then, after some algebra using the dressed states computed above,
the unitary evolution operator (\ref{eq:ua}) is given by,
\begin{eqnarray}
    U_0(t)&=&
    e^{i\frac{\Delta}{2}t-ig\sqrt{n+1}t}|a,n,t\rangle\langle a,n,0|+
    e^{-i\frac{\Delta}{2}t+ig\sqrt{n+1}t}|b,n,t\rangle\langle b,n,0|
    \nonumber \\
        &=&\cos\left(\frac{{\cal R}_n}{2}t\right)
    (e^{-i\frac{\Delta}{2}t}|1,n+1\rangle\langle 1,n+1| +
     e^{i\frac{\Delta}{2}t}|2,n\rangle\langle 2,n|) \label{eq:u0} \\
     &-&i\sin\left(\frac{{\cal R}_n}{2}t\right)
     (e^{-i\frac{\Delta}{2}t}|1,n+1\rangle\langle 2,n| +
      e^{i\frac{\Delta}{2}t}|2,n\rangle\langle 1,n+1|) \nonumber
\end{eqnarray}
that, for $|\psi(0)\rangle=c_{1,n+1}(0)|1,n+1\rangle
+c_{2,n}(0)|2,n\rangle$, gives
\begin{eqnarray}
    |\psi(t)\rangle_I &\approx&
    \left[\cos\left(\frac{{\cal R}_n}{2}t\right)c_{1,n+1}(0)
     -i\sin\left(\frac{{\cal R}_n}{2}t\right)c_{2,n}(0)\right]
     e^{-i\frac{\Delta}{2}t}|1,n+1\rangle \nonumber \\
    & & \\
    &+&\left[\cos\left(\frac{{\cal R}_n}{2}t\right)c_{2,n}(0)
     -i\sin\left(\frac{{\cal R}_n}{2}t\right)c_{1,n+1}(0)\right]
     e^{i\frac{\Delta}{2}t}|2,n\rangle \nonumber
\end{eqnarray}
that is the exact form of eqs.(\ref{eq:sjcb}) when higher
order terms beyond the leading one are neglected, i.e. when
$\lambda\rightarrow\infty$, as expected from the results of
Ref.\cite{fra1}.

In order to go to higher orders, we have to compute $H'(t)$
from eq.(\ref{eq:h1}). Again, using the above expressions
for the dressed states one gets
\begin{eqnarray}
    H'(t)&=&-\frac{\Delta}{2}
    \left[\cos\left({\cal R}_n t\right)
    (|1,n+1\rangle\langle 1,n+1| -
     |2,n\rangle\langle 2,n|)\right. \nonumber \\
     & & \\
     &-&\left.i\sin\left({\cal R}_n t\right)
     (|1,n+1\rangle\langle 2,n| -
      |2,n\rangle\langle 1,n+1|)\right] \nonumber
\end{eqnarray}
so that, the first order
correction to the leading order evolution operator
$U_0(t)$ of eq.(\ref{eq:u0}) is given by
\begin{eqnarray}
    U_1(t)=-iU_0(t)\int_0^t dt_1H'(t_1)=
    i\frac{1}{\lambda}\sin\left(\frac{{\cal R}_n}{2}t\right)
    (e^{-i\frac{\Delta}{2}t}|1,n+1\rangle\langle 1,n+1|-
     e^{i\frac{\Delta}{2}t}|2,n\rangle\langle 2,n|)
\end{eqnarray}
that gives the first order correction
\begin{eqnarray}
    |\delta_1\psi(t)\rangle_I=
    i\frac{1}{\lambda}\sin\left(\frac{{\cal R}_n}{2}t\right)
    (e^{-i\frac{\Delta}{2}t}c_{1,n+1}(0)|1,n+1\rangle-
     e^{i\frac{\Delta}{2}t}c_{2,n}(0)|2,n\rangle)
\end{eqnarray}
again in agreement with the Taylor expansion as given
in eqs.(\ref{eq:sjcb}), to order $\frac{1}{\lambda}$.
So, in the same way we have at the second order
\begin{eqnarray}
    U_2(t)&=&-U_0(t)\int_0^t dt_1 H'(t_1)
    \int_0^{t_1}dt_2 H'(t_2)= \nonumber \\
    & &i\frac{1}{2\lambda^2}
    \left\{\left[\sin\left(\frac{{\cal R}_n}{2}t\right)
    -\frac{{\cal R}_n}{2}t
    \cos\left(\frac{{\cal R}_n}{2}t\right)\right]
    (e^{i\frac{\Delta}{2}t}|2,n \rangle\langle 1,n+1|+
    e^{-i\frac{\Delta}{2}t}|1,n+1 \rangle\langle 2,n|)\right. \\
    &-&\left.i\frac{{\cal R}_n}{2}t
    \sin\left(\frac{{\cal R}_n}{2}t\right)
    (e^{-i\frac{\Delta}{2}t}|1,n+1\rangle\langle 1,n+1|+
     e^{i\frac{\Delta}{2}t}|2,n\rangle\langle 2,n|)\right\} \nonumber
\end{eqnarray}
then, one has
\begin{eqnarray}
    |\delta_2\psi(t)\rangle_I&=&
    i\frac{1}{2\lambda^2}
    \left\{\left[\sin\left(\frac{{\cal R}_n}{2}t\right)
    -\frac{{\cal R}_n}{2}t
    \cos\left(\frac{{\cal R}_n}{2}t\right)\right]
    (e^{i\frac{\Delta}{2}t}c_{1,n+1}(0)|2,n\rangle+
     e^{-i\frac{\Delta}{2}t}c_{2,n}(0)|1,n+1\rangle)\right.
     \nonumber \\
     & & \\
    &-&i\left.\frac{{\cal R}_n}{2}t
    \sin\left(\frac{{\cal R}_n}{2}t\right)
    (e^{-i\frac{\Delta}{2}t}c_{1,n+1}(0)|1,n+1\rangle+
     e^{i\frac{\Delta}{2}t}c_{2,n}(0)|2,n\rangle)\right\}. \nonumber
\end{eqnarray}
The agreement with the Taylor expansion as given
in eqs.(\ref{eq:sjcb}), to order $\frac{1}{\lambda^2}$,
is complete.

As expected from results of Ref.\cite{fra1}, the adiabatic
approximation and its higher order corrections turn out
to be a nice method for asympotic analysis of the
Schr\"{o}dinger equation, being the dual of the well-known
Dyson series and explaining in this way the nice
working of the method of dressed states currently used
in quantum optics. No slowly varying of the parameters of the
Hamiltonian is involved as one could expect for
the adiabatic approximation.

\section{Perturbative analysis of models for high-order harmonic generation}
\subsection{Models}

Several models are
currently used to account for high-order harmonic generation.
The first model considered \cite{ebe} has been a
one-dimensional model described by the Schr\"{o}dinger equation
\begin{equation}
    \left[-\frac{1}{2}\frac{\partial^2}{\partial x^2}+V(x)
    -x\epsilon_0(t)\sin{\omega_L t}\right]\Psi(x,t)=
    i\frac{\partial\Psi(x,t)}{\partial t} \label{eq:tdse}
\end{equation}
being $\epsilon_0(t)$ a function taking in account the time
to rise the laser field to its maximum value, $\omega_L$ the
frequency of the laser field and $V(x)$ a simple representative
binding potential for the atom. A choice currently found in
literature is $V(x)=-\frac{1}{\sqrt{1+x^2}}$.
Beside numerical methods that are very computer demanding,
other methods as Floquet theory
have also been applied \cite{Potv} for the full three-dimensional case.
A fruitful understanding of harmonic generation through semiclassical
ideas has also been yielded in Ref.\cite{Cork}. By these semiclassical
results, a non-perturbative quantum model has been obtained
\cite{Huil}. Beside, a first approach by second quantization
has also been given where a hint was put forward that
harmonic generation is a first order effect \cite{Cras}.
Analytical expression are barely given as
all these models have been solved or
numerically either non-perturbatively
so to require at some step numerical computation.
Another model is a simpler two-level system
described by the hamiltonian \cite{mil,kei,maq}
\begin{equation}
    H=\frac{\omega_0}{2}(|2\rangle\langle 2|-|1\rangle\langle 1|)
    -x\epsilon_0(t)\left\{
    \begin{array}{c}
    \sin{\omega_L t} \\ \cos{\omega_L t}
    \end{array}                \label{eq:tls}
    \right\}
\end{equation}
and
\begin{equation}
    x=-d_{12}(|1\rangle\langle 2|+|2\rangle\langle 1|)
\end{equation}
being $d_{12}$ the matrix element of the atomic dipole. This model
is well-known in quantum mechanics. A first hint to a strong coupling
perturbative solution was given by Meystre \cite{mey} that used it
to describe an atom inside a Fabry-Perot cavity.
The series till first order and the way to compute higher orders
for strong coupling were finally obtained in Ref.\cite{rg} where
it was shown that a shift of the levels of the atom occurs.

Indeed, this two-level model seems very effective in
describing high-order harmonic generation too. The two
physical situations of a Fabry-Perot cavity strongly
coupled with an atom and an atom in a strong laser
field seems described by the same hamiltonian. But
this should not come out as a surprise. What really
matters here is the existence of the shift of the
energy levels of the atom in these situations that,
for the case of high-order harmonic generation can
change the spectrum of hyper-Raman lines and so,
can be measured experimentally.

Beside, as we are going to show, the leading order
solution found by Meystre is not enough to get the
power spectrum computed through the Fourier
transform of the equation
\begin{equation}
    x(t)=\langle\Psi(t)|x|\Psi(t)\rangle \label{eq:x}
\end{equation}
In fact, by the dual Dyson series one can see that
high-order harmonic generation
is actually a first order effect. In this way we are
able to reproduce the results obtained in Ref.\cite{maq}
by Floquet method, but having an analytical expression
to be compared with experiments. As a by-product we have
that the hyper-Raman lines can be shifted. Through this approach
the computation can be pushed to any order, coping always
with definite analytical expressions.

The model (\ref{eq:tdse}) can also be treated by this approach.
Indeed, an application to multiphoton ionization has been found
by Salamin \cite{sal}. The leading order solution should be
written as
\begin{equation}
    \psi(x,t)\approx
    e^{ix\int_0^tdt'\epsilon_0(t')\sin{\omega_L t'}}\phi_n(x)
    \label{eq:lead0}
\end{equation}
being
\begin{equation}
    \left[-\frac{1}{2}\frac{\partial^2}{\partial x^2}+V(x)
    \right]\phi_n(x)=E_n\phi_n(x).
\end{equation}
It easy to see that probability transitions given by
$w_{mn}(t)=\int_{-\infty}^{+\infty}dx\phi_m(x)\psi(x,t)$
are not trivial and can be computed also for the continuos
part of the spectrum. But, as we are going to show using
the two-level model and as can be seen by the look
of the leading order solution (\ref{eq:lead0}),
we need to compute the first order correction to it
to account for high-order harmonic generation. We
do not pursue the study of this model further here, as
the two-level model can give a satisfactory account of
all this matter in a simpler way. We just note that
in this way, more complex models than
that of eq.(\ref{eq:tdse}), through perturbation methods,
could be taken into account.

\subsection{Perturbative analysis for high-order harmonic generation}

To fix the ideas, we consider the two-level model of
Ref.\cite{maq}, that is, eq.(\ref{eq:tls}) with a
cosine perturbation. Dyson series using probability
amplitudes and its dual solution to first order
of this model through operatorial methods were given
in Ref.\cite{rg}. So, we avoid the analysis by
the Dyson series of this model discussed in depth
in \cite{rg} and Refs. therein. Instead, we use the dual
Dyson series to show that it is equivalent to
the operatorial method used in \cite{rg} and
presented initially in Ref.\cite{fra2}.

The rising of the laser field accounted for by the
function $\epsilon_0(t)$ is taken as istantaneous
to make the computations simpler,
that is, we take $\epsilon_0(t)=\Omega=$constant.

In interaction picture, the Hamiltonian (\ref{eq:tls}) is
given by
\begin{equation}
    H_I=\Omega d_{12}\cos{\omega_L t}(e^{-i\omega_0 t}
    |1\rangle\langle 2|+e^{i\omega_0 t}|2\rangle\langle 1|)
\end{equation}
Then, computing the dual Dyson series,
for the eigenvalue $\Omega d_{12}\cos{\omega_L t}$ we get
the eigenvector
\begin{equation}
    |b,t\rangle=\frac{1}{\sqrt{2}}(e^{i\omega_0 t}|2\rangle+|1\rangle)
\end{equation}
and for the eigenvalue $-\Omega d_{12}\cos{\omega_L t}$
the eigenvector
\begin{equation}
    |a,t\rangle=\frac{1}{\sqrt{2}}(|2\rangle-e^{-i\omega_0 t}|1\rangle).
\end{equation}
These are the dressed states for this model. The corresponding Berry
phases are given by
\begin{eqnarray}
    \dot{\gamma}_b(t)&=&\frac{\omega_0}{2} \nonumber \\
    & & \nonumber \\
    \dot{\gamma}_a(t)&=&-\frac{\omega_0}{2}. \nonumber \\
\end{eqnarray}
It is interesting to note here, that Berry phases originate from
the energies of the levels of the unpertubed atom.

All this gives the unitary evolution
\begin{equation}
    U_0(t)=e^{-i\frac{\omega_0}{2}t}
    e^{i\frac{\Omega d_{12}}{\omega_L}\sin{\omega_L t}}
    |a,t\rangle\langle a,0|+
    e^{i\frac{\omega_0}{2}t}
    e^{-i\frac{\Omega d_{12}}{\omega_L}\sin{\omega_L t}}
    |b,t\rangle\langle b,0|
\end{equation}
that yields in terms of the bare states $|1\rangle$ and $|2\rangle$
\begin{equation}
    U_0(t)=
    \cos\left(\frac{\Omega d_{12}}{\omega_L}\sin{\omega_L t}\right)
    (e^{-i\frac{\omega_0}{2}t}|1\rangle\langle 1|+
    e^{i\frac{\omega_0}{2}t}|2\rangle\langle 2|)
    -i\sin\left(\frac{\Omega d_{12}}{\omega_L}\sin{\omega_L t}\right)
    (e^{-i\frac{\omega_0}{2}t}|1\rangle\langle 2|+
    e^{i\frac{\omega_0}{2}t}|2\rangle\langle 1|).
\end{equation}
We can reformulate the above operator as a matrix by taking
for the bare states
\begin{eqnarray}
    |1\rangle =
    \left(\begin{array}{cc} 0 \\ 1 \end{array} \right), & \;\; &
    |2\rangle =
    \left(\begin{array}{cc} 1 \\ 0 \end{array} \right)
\end{eqnarray}
so to have
\begin{eqnarray}
    U_0(t)=\left(\begin{array}{cc}
    e^{i\frac{\omega_0}{2}t}
    \cos\left(\frac{\Omega d_{12}}{\omega_L}\sin{\omega_L t}\right) &
    -ie^{i\frac{\omega_0}{2}t}
    \sin\left(\frac{\Omega d_{12}}{\omega_L}\sin{\omega_L t}\right) \\
    -ie^{-i\frac{\omega_0}{2}t}
    \sin\left(\frac{\Omega d_{12}}{\omega_L}\sin{\omega_L t}\right) &
    e^{-i\frac{\omega_0}{2}t}
    \cos\left(\frac{\Omega d_{12}}{\omega_L}\sin{\omega_L t}\right)
    \end{array}\right)
\end{eqnarray}
It is not difficult to see that the above operator can be rewritten
through the Pauli matrices $\sigma_1,\sigma_2,\sigma_3$ as
\begin{equation}
    U_0(t)=e^{i\frac{\omega_0}{2}\sigma_3t}
           e^{-i\sigma_1
           \frac{\Omega d_{12}}{\omega_L}\sin{\omega_L t}}
\end{equation}
then, by eliminating the prefactor due to interaction picture, we are left
with the leading order result of Ref.\cite{rg} for the wave function
\begin{equation}
    |\Psi(t)\rangle\approx e^{-i\sigma_1
           \frac{\Omega d_{12}}{\omega_L}\sin{\omega_L t}}
           |\Psi(0)\rangle. \label{eq:or0}
\end{equation}
In the same way, we can compute higher order corrections to the above by
computing $H'(t)$ for the dual Dyson series. In the bare states, using
again the dressed ones, one has
\begin{equation}
    H'(t)=\frac{\omega_0}{2}\left[
    \cos\left(2\frac{\Omega d_{12}}{\omega_L}\sin{\omega_L t}\right)
   (|2\rangle\langle 2|-|1\rangle\langle 1|)
    -i\sin\left(\frac{\Omega d_{12}}{\omega_L}\sin{\omega_L t}\right)
    (|2\rangle\langle 1|-|1\rangle\langle 2|)\right]
\end{equation}
that is,
\begin{equation}
    H'(t)= \frac{\omega_0}{2}
           e^{i\sigma_1
           \frac{\Omega d_{12}}{\omega_L}\sin{\omega_L t}}
           \sigma_3
           e^{-i\sigma_1
           \frac{\Omega d_{12}}{\omega_L}\sin{\omega_L t}}
\end{equation}
in agreement with the computation of the first order correction
computed through operatorial methods in Ref.\cite{rg}. The two
series are identical as it should be expected.

So, the solution for the system (\ref{eq:tls}) till first order can
be written as \cite{rg}
\begin{eqnarray}
    |\Psi(t)\rangle&=&e^{-i\sigma_1
           \frac{\Omega d_{12}}{\omega_L}\sin{\omega_L t}}
           \times \nonumber \\
           &\left[\right.& I-i\frac{\omega_0}{2}
           J_0\left(\frac{2\Omega d_{12}}{\omega_L}\right)t\sigma_3
           \nonumber \\
           &-&i\omega_0\sum_{n=1}^{\infty}
           J_{2n}\left(\frac{2\Omega d_{12}}{\omega_L}\right)
           \frac{\sin(2n\omega_L t)}{2n\omega_L}\sigma_3 \nonumber \\
           &+&i\omega_0\sum_{n=0}^{\infty}
           J_{2n+1}\left(\frac{2\Omega d_{12}}{\omega_L}\right)
           \frac{\cos((2n+1)\omega_L t)-1}{(2n+1)\omega_L}
           \sigma_2           \nonumber \\
           &+&\left.\cdots\right]|\Psi(0)\rangle. \label{eq:fo}
\end{eqnarray}
where use has been made of the operatorial identity
\begin{eqnarray}
    e^{\pm i\sigma_k z\sin{\phi}}=
           J_0(z)+2\sum_{n=1}^{\infty}J_{2n}(z)\cos(2n\phi)
           \pm 2i\sigma_k \sum_{n=0}^{\infty}J_{2n+1}(z)
           \sin((2n+1)\phi)
\end{eqnarray}
with $\sigma_k$ one of the Pauli matrices and $J_n(z)$ Bessel functions
of integer order. The secular term in eq.(\ref{eq:fo}) can be resummed
away by renormalization group methods, as shown in Ref.\cite{rg}, giving
the renormalized levels of the atom in the laser field. Then, the
solution one has to use to compute the power spectrum is
\begin{eqnarray}
    |\Psi(t)\rangle &=&e^{-i\sigma_1
           \frac{\Omega d_{12}}{\omega_L}\sin{\omega_L t}}
           \times \nonumber \\
           &\left[\right.&I-i\omega_0\sum_{n=1}^{\infty}
           J_{2n}\left(\frac{2\Omega d_{12}}{\omega_L}\right)
           \frac{\sin(2n\omega_L t)}{2n\omega_L}\sigma_3 \nonumber \\
           &+&i\omega_0\sum_{n=0}^{\infty}
           J_{2n+1}\left(\frac{2\Omega d_{12}}{\omega_L}\right)
           \frac{\cos((2n+1)\omega_L t)-1}{(2n+1)\omega_L}
           \sigma_2           \nonumber \\
           &+&\left.\cdots\right]e^{
           -i\frac{\omega_0}{2}
           J_0\left(\frac{2\Omega d_{12}}{\omega_L}\right)t\sigma_3}
           |\Psi(0)\rangle. \label{eq:for}
\end{eqnarray}

It easy to see that if we just limit our analysis to eq.(\ref{eq:or0}),
we are not able to obtain the spectrum of the harmonics.
In fact, one would have from eq.(\ref{eq:x})
$x(t)=\langle\Psi(t)|x|\Psi(t)\rangle=
-d_{12}\langle\Psi(0)|\sigma_1|\Psi(0)\rangle=$constant. Instead,
using eq.(\ref{eq:for}) one has at first order
\begin{eqnarray}
    x(t)&=&-d_{12}\left[
    c_2c_1^*e^{-i\omega_{0R}t}+c_2^*c_1e^{i\omega_{0R}t}\right. \nonumber \\
        &+&(|c_1|^2-|c_2|^2)\omega_0
           \sum_{n=0}^{\infty}
           J_{2n+1}\left(\frac{2\Omega d_{12}}{\omega_L}\right)
           \frac{\cos((2n+1)\omega_L t)-1}
           {(n+\frac{1}{2})\omega_L} \nonumber \\
       &+&\left.i(c_2^*c_1e^{i\omega_{0R}t}-c_2c_1^*e^{-i\omega_{0R}t})
           \omega_0\sum_{n=1}^{\infty}
           J_{2n}\left(\frac{2\Omega d_{12}}{\omega_L}\right)
           \frac{\sin(2n\omega_L t)}{n\omega_L}\right]
\end{eqnarray}
being $\omega_{0R}=\omega_0J_0\left(\frac{2\Omega d_{12}}{\omega_L}\right)$
the renormalized separation of the two levels of the atom and introducing the
population distribution $c_1$ and $c_2$ for the bare levels of the atom
through the initial state $|\Psi(0)\rangle$. This is exactly
the form one must have for the high-order harmonic generation using
this model, as showed in Ref.\cite{maq} by the Floquet method. In fact,
we have odd harmonics of intensity
$(|c_1|^2-|c_2|^2)^2$, while the latter term is due to the hyper-Raman lines
at $\omega_{0R}\pm 2n\omega_L$ of intensity $|c_2|^2|c_1|^2$. But now, an
explicit analytic expression for the power spectrum is given, so that, a
clear understanding of all the parameters involved into atom and strong
laser field interaction is obtained. Particularly, we have an exact
expression for the shift of the levels of the atom that now we know how
to measure: If we take initially $|c_1|=|c_2|$ we will be just left
with hyper-Raman lines into the spectrum. These lines can be shifted
by varying the ratio $\frac{2\Omega d_{12}}{\omega_L}$, given by the
parameters to be controlled into the experiment. By observing how
these lines move we can obtain a measure of the shifts of the atom levels.
By the expression of $x(t)$, it is clear that the only lines that can be
moved are indeed the hyper-Raman lines.

The advantges above the Floquet method used in Ref.\cite{maq}
are evident as we have an explicit analytical formula for the
spectrum with all the functional dependencies on the parameters
entering into the model explicitly expressed.

\section{Conclusions}

The way dressed states are currently treated in quantum optics
becomes to compute the eigenstates and eigenvalues pertaining or the
interaction term either the full Hamiltonian of a given system.
In this paper we have shown how the dual Dyson series can
give both an understanding of this approach and a tool to improve it.
Indeed, the existence of a dual Dyson series
permits, as we have shown, a perturbative study of
atom-field interaction in different regions of the parameter
space of a given model. The case of the Jaynes-Cummings model,
fully exploited in this paper, is examplary in this sense. Beside,
the dual Dyson series is nothing else than the quantum adiabatic
approximation and its higher order corrections that in this way
prove to be a very powerful tool to obtain asymptotic approximations
to the Schr\"{o}dinger equation.

The theory is applied to the analysis of models for high-order
harmonic generation giving an explicit expression for the
power spectrum. In this way new experiments can be thought where,
properly changing the parameters involved, measures of the energy
levels of the atom in interaction in a strong laser field could
be accomplished. Beside, a control on the amplitudes of the
harmonics and the position of the cut-off on the spectrum could
be obtained.

\end{document}